# DISCUSSION OF "2004 IMS MEDALLION LECTURE: LOCAL RADEMACHER COMPLEXITIES AND ORACLE INEQUALITIES IN RISK MINIMIZATION" BY V. KOLTCHINSKII

BY XIAOTONG SHEN AND LIFENG WANG

*University of Minnesota*

Koltchinskii is to be congratulated for developing a unified framework. This elegant framework is general and allows a user to apply it directly instead of deriving bounds in each risk minimization problem. In the past decade, the problem of risk minimization has been extensively studied in function estimation and classification. In function estimation, it has been investigated using the empirical process technique under the name of minimum contrast or sieve estimation in, for instance, [2, 3, 7, 12, 13, 15]. In classification, it has been studied in a similar fashion; cf. [1, 6, 10, 11]. The general framework derived in this article yields an upper bound of the excess risk through local Rademacher complexities. When applying such a framework to a specific problem, attention is necessary with regard to the specific problem structure that may matter greatly.

Our discussion will be centered in two aspects: (1) the role that the variance and mean play, particularly in classification, and (2) practicability of an empirical complexity.

## 1. The role of variance and mean.

1.1. *Variance–mean relationship and the margin condition.* As noted in the paper, one key idea to recover the optimal rate of convergence is to bound the local complexity $E\sup_{f \in \mathcal{F}: P(f-\bar{f}) \le \delta} |(P_n - P)(f - \bar{f})|$ instead of the global one. This is achieved by bounding $Var(f - \bar{f})$, or sufficiently the second moment $P(f - \bar{f})^2$, by the mean $P(f - \bar{f})$. Such a variance–mean relationship was essentially used in [7] in a slightly more general form of

(1) $$Var(f(X) - \bar{f}(X)) \le a[E(f(X) - \bar{f}(X))]^{2\beta}$$

for some constants $a > 0$ and $\beta > 0$, on which an iterative improvement approach is employed to derive fast rates of convergence by exploring $\sup_A |(P_n - P)(f - \bar{f})|$ over local set $A$. This is analogous to the fixed point approach used in the present paper.

In what follows, we argue that (1) is more fundamental than the popular margin (low noise) assumption (cf. [10]) commonly used in classification, resulting in fast rates of convergence. In classification, (1) summarizes not only the local behavior of the optimal decision function near its classification boundary but also the global behavior of the optimal decision function, whereas the margin assumption describes only the former. To be more specific, consider Tsybakov's classification example as discussed in Section 6.1 of the present paper. Following the notations of [10], denote $Z$ to be the input/output pair $(X, Y)$ with $X \in \mathbf{R}^d$ and $Y = 0, 1$. Let $\mathcal{G}^* = \{G \subset \mathbf{R}^d\}$ be a class of classification sets. Now define $f = f_G$ to be $I(Y \neq I(X \in G))$, which is the 0–1 classification error loss for set $G \in \mathcal{G}^*$ given $Z$. Easily, it can be seen that the margin assumption (A1) in [10] implies (1) with $\beta = \frac{1}{2\kappa}$, and assumption (A2) there implies the $L_2(P)$ bracketing entropy of $\mathcal{F} = \{f_G : G \in \mathcal{G}^*\}$ to be of order $\varepsilon^{-2\rho}$ with $0 < \rho < 1$. Then an application of Theorem 2 of Shen and Wong [7] with $\alpha = 1/2$, $\beta = 1/(2\kappa)$ and $r = 2\rho$ yields a rate of convergence of the excess risk $\rho(f_{\hat{G}_n}, f_{G^*})$, or the Bayesian regret, $n^{-1/(4\alpha - \min(\alpha, \beta)(2-r))} = n^{-\kappa/(2\kappa + \rho - 1)}$, which agrees with the result of Tsybakov [10] and the present paper. Here $\rho(f_G, f_{G^*}) = \rho(f, \bar{f}) = E(f_G(Z) - f_{G^*}(Z))$ with $\bar{f} = f_{G^*}(Z)$ defined by the optimal classification set $G^*$. This example indicates that (1) is actually weaker than the margin assumption (A1). Furthermore, (1) continues to be a key assumption for risk minimization in classification even when (A1) breaks down, as in linear SVM with the hinge loss. This is because in this case $\bar{f}$ no longer approximates the Bayes rule in the sense of [10].

1.2. *Variance–mean relationship in margin-based classification.* Consider an equivalent version of (1) in regression and classification:

$$(2) \quad Var(l(Y, f(X)) - l(Y, \bar{f}(X))) \leq a[E(l(Y, f(X)) - l(Y, \bar{f}(X)))]^{2\beta},$$

where $(X, Y)$ is an observation pair, $l$ is a loss function and $f$ is a parameter in $\mathcal{F}$.

The present paper nicely illustrates importance of the variance–mean relationship in least squares regression in which $\beta = 1/2$ in (2). In classification, however, the situation is much more complex. As illustrated in [14], (2) may depend on the choice of loss functions and $\mathcal{F}$. For simplicity, we assume $y = \pm 1$ as opposed to $0, 1$ in what follows. For $\psi$-learning [8], $l(y, f(x)) = I[yf(x) < 0] + (1 - yf(x))I[0 \leq yf(x) < 1]$, (2) holds with $\beta = 1/(2\kappa)$, where $\kappa$ is the exponent given assumption (A1). As a result, the aforementioned fast rate $n^{-\kappa/(2\kappa+\rho-1)}$ in Section 1.1 can be realized by $\psi$-learning, provided that the bracketing $L_2$ entropy of the class of candidate classification sets induced by decision functions is of order $\varepsilon^{-2\rho}$ for $0 < \rho < 1$. For SVM, $l(y, f(x)) = [1 - yf(x)]_+$ and hence (2) is met with $\beta = 1/2$ generally for a finite-dimensional linear space $\mathcal{F}$. However, when $\mathcal{F}$ is sufficiently rich, $\beta = 1/2$ for the separable case but is essentially 0 for the nonseparable case.

1.3. *Variance–mean relationship when the candidate function class $\mathcal{F}$ is large.* Another phenomenon worthwhile mentioning is that (1) may not be useful for improving the excess risk bound when $\mathcal{F}$ is very large. For instance, when the metric entropy of $\mathcal{F}$ is of the order $\varepsilon^{-2\rho}$ with $\rho = 1$, a rate $n^{-1/2} \log n$ can be realized (Theorem 2 of [7]), where $\beta$ in (1) does not enter into the rate expression. This is in contrast to the fast rate obtained in Section 1.2. A situation like this occurs in the $L_1$-norm SVM variable selection with the number of candidate variables $d$ greatly exceeding that of the sample size $n$, where $\mathcal{F} = \{f = \theta^T x : \|\theta\|_1 \leq s, \theta \in \mathbf{R}^d\}$ with a tuning parameter $s > 0$, and the entropy of $\mathcal{F}$ is of order of $(\log d)\varepsilon^{-2}$. This leads to a rate of $(n^{-1} \log d)^{1/2} \log(n(\log d)^{-1})$ (cf. [14]), as long as $d$ grows no faster than $\exp(n)$.

**2. Empirical complexity as a way of model selection.** With regard to model selection, the author advocates a model selection criterion through penalization that mimics an oracle inequality of some type (Section 5), which is an upper bound of the excess risk. The selection criterion, or a data-dependent upper bound, estimates the oracle inequality. For a model selection criterion constructed in this manner, several important issues remain. First, an oracle (upper) inequality through a concentration inequality could be rough in the sense that the difference between the upper bound and the actual excess risk is large, although it dramatically simplifies the process of estimating the excess risk. Consequently, the optimal model selected by the model selection criterion may be inaccurate due to the bias introduced by an imprecise upper bound of the excess risk, particularly in the finite-sample situation. This phenomenon has been noted in [5] when AIC and BIC were compared against Vapnik's structural risk minimization via a penalty based on the VC-dimension, with respect to the accuracy of prediction. Second, it appears rather difficult to track the constants in the penalties theoretically and empirically. Theoretically, $\tilde{\pi}_n(k)$ and $\tilde{K}$ in the oracle inequality may be imprecise in that many "numerical constants" are suited for the purpose. Then can these terms be optimally determined? Empirically, it seems unnecessary that $\tilde{\pi}_n(k)$ and $\tilde{K}$ are precisely estimated by $\hat{\pi}(k)$ and $\hat{K}$; even an inconsistent estimator can give the desired result. While a rate of convergence result is useful in providing an insight into the problem of model selection, estimation of an overly simplified oracle inequality may not be precise enough to determine all required constants—further developments may be needed.

The foregoing discussion brings up an interesting and important point: how to balance mathematical tractableness and the accuracy of risk estimation. We now turn our attention to covariance penalties in the framework of model selection via penalization, which directly estimates the risk/loss based on optimal predictive estimation. Covariance penalties that are approximately unbiased for estimating the risk are shown to be more precise in prediction than their competitor cross-validation in [4]. A general construction of covariance penalties can be found in [4, 9] in a family of Q-error losses including the Kullback–Leibler loss and the 0–1 classification error loss. General estimation methods for covariance penalties include bootstrap and data perturbation. In contrast, covariance penalties are less tractable theoretically than the penalties in an oracle inequality.

School of Statistics  
University of Minnesota  
224 Church Street Southeast  
Minneapolis, Minnesota 55455  
USA  
E-mail: xshen@stat.umn.edu